\documentclass[english,aps,prl,reprint,superscriptaddress,nolongbibliography]{revtex4-2}
\usepackage[T1]{fontenc}
\usepackage[latin9]{inputenc}
\usepackage{amsmath}
\usepackage{amssymb}
\usepackage{graphicx}
\usepackage{wasysym}
\usepackage{color}
\usepackage{physics}

\makeatletter
\@ifundefined{textcolor}{}
{%
 \definecolor{BLACK}{gray}{0}
 \definecolor{WHITE}{gray}{1}
 \definecolor{RED}{rgb}{1,0,0}
 \definecolor{GREEN}{rgb}{0,1,0}
 \definecolor{BLUE}{rgb}{0,0,1}
 \definecolor{CYAN}{cmyk}{1,0,0,0}
 \definecolor{MAGENTA}{cmyk}{0,1,0,0}
 \definecolor{YELLOW}{cmyk}{0,0,1,0}
 \definecolor{figa}{rgb}{0.7143, 0.7143, 0.7143}
 \definecolor{figb}{rgb}{0.4416, 0.7490, 0.4322}
 \definecolor{figc}{rgb}{0.3639, 0.5755, 0.7484}
 \definecolor{figd}{rgb}{0.9153, 0.2816, 0.2878}
 \definecolor{fige}{rgb}{1.0000, 0.5984, 0.2000}
 \definecolor{figf}{rgb}{0.6365, 0.3753, 0.6753}
}

\usepackage{pslatex}

\usepackage{babel}

\makeatother

\begin{document}

\title{High Compression Blue-Detuned Magneto-Optical Trap of Polyatomic Molecules}

\author{Christian Hallas}
\email{christianhallas@g.harvard.edu}

\author{Grace K. Li}
\author{Nathaniel B. Vilas}
\author{Paige Robichaud}
\author{Lo\"{i}c Anderegg}
\author{John M. Doyle}

\affiliation{Department of Physics, Harvard University, Cambridge, MA 02138, USA}
\affiliation{Harvard-MIT Center for Ultracold Atoms, Cambridge, MA 02138, USA}

\date{\today}

\begin{abstract}
We demonstrate a blue-detuned magneto-optical trap (MOT) of a polyatomic molecule, calcium monohydroxide (CaOH). We identify a novel MOT frequency configuration that produces high spatial compression of the molecular cloud. This high compression MOT achieves a cloud radius of $59(5)~\text{\textmu m}$ and a peak density of $8(2) \times 10^8~\text{cm}^{-3}$, the highest reported density for a molecular MOT to date. 
We compare our experimental studies of blue-detuned MOTs for CaOH and compare with Monte-Carlo simulations, finding good agreement.
\end{abstract}

\maketitle
Ultracold molecules are novel quantum systems with rich internal level structures and large intrinsic electric dipole moments, which makes them promising platforms for pursuing a wide range of scientific applications. These include quantum information processing~\cite{demille2002quantum, yelin2006schemes, ni2018dipolar, sawant2020ultracold, wei2011entanglement, yu2019scalable, albert2020robust}, quantum simulation~\cite{micheli2006toolbox, gorshkov2011quantum, wall2014quantum}, searches for physics beyond the Standard Model~\cite{kozyryev2017precision, kozyryev2021enhanced, norrgard2019nuclear}, and studies of quantum chemistry~\cite{heazlewood2021towards} and collisions~\cite{cheuk_2020}. 
%Motivated by their promise, significant experimental effort has been made in extending laser cooling techniques to molecules. 
Over the last decade, significant experimental effort has been made in developing laser cooling and trapping techniques for molecules. Magneto-optical traps (MOTs) have been realized for a number of molecular species and, as with atoms, the MOT has become an important tool for rapid spatial compression and cooling of molecules to submillikelvin temperatures~\cite{barry2014magneto, anderegg2017radio, collopy20183d, CaOH_MOT}. In combination with sub-Doppler laser cooling techniques~\cite{deep_laser_cooling, SrF_cooling, YO_cooling, CaOH_MOT}, the development of molecular MOTs has enabled the loading of laser-cooled molecules into conservative optical traps~\cite{Anderegg2018, CaOH_ODT, SrF_ODT}. Recently, tweezer arrays of individual molecules, including polyatomic molecules, have also been demonstrated with laser-cooled molecules~\cite{anderegg2019science, CaF2_tweezer, vilas2024tweezers}. Several milestones towards the above applications have been realized, e.g., studies of molecular collisions at ultracold temperatures~\cite{cheuk_2020, anderegg2021observation}, the demonstration of entanglement between individual molecules~\cite{CaF2_tweezer, CaF_dipolar}, and sideband cooling to the motional ground state~\cite{CaF2_sideband, CaF_sideband}. 

Conventional molecular MOTs rely on Doppler cooling and trapping forces achived by optically cycling on an electronic transition using ``red-detuned'' MOT light, i.e., the MOT light frequency is smaller than the resonance frequency of the transition. Rotationally closed cycling is accomplished by driving so-called type-two transitions with $J \geq J'$, where $J$ ($J'$) is the quantum number for the total angular momentum of the ground (excited) electronic state~\cite{stuhl2008magneto}. At low molecular velocities, type-two systems exhibit strong Sisyphus-like forces, which for red-detuned light give rise to heating~\cite{devlin2016three}. 
Compared with typical atomic MOTs, which drive type-one transitions with $J < J'$, red-detuned molecular MOTs (``red MOTs'') therefore have relatively high temperatures ($\sim$$1$~\text{mK}) and low spatial confinement (cloud sizes of $\sim$$1$~\text{mm})~\cite{barry2014magneto, anderegg2017radio, collopy20183d, CaOH_MOT}.
The weak compression of red molecular MOTs has limited the loading of molecules into conservative optical traps to a few percent.
Higher molecular number densities are desirable for a number of scientific goals and, in particular, for any experiments involving collisions, including evaporative cooling to quantum degeneracy.
With an eye towards improving molecular MOTs, blue-detuned MOTs were proposed~\cite{tarbutt2015magneto} and demonstrated using Rb atoms~\cite{Rb} by Tarbutt and coworkers. With blue-detuned MOT light, heating turns into Sisyphus-like cooling for molecules at low velocities. Recently, blue-detuned MOTs (``blue MOTs'') were demonstrated for diatomic molecules~\cite{YO, jorapur2023, CaF} with great success. 
Blue MOTs realized with diatomic molecules have achieved both lower temperatures and higher densities compared to red-detuned molecular MOTs. 

In this Letter, we demonstrate a blue MOT for a polyatomic molecule, calcium monohydroxide (CaOH), and identify a MOT light frequency scheme for CaOH that realizes higher spatial compression than existing blue molecular MOTs.
In this ``1+2'' scheme, a single-frequency light beam addresses one resolved ground state and an additional pair of light beams address a different resolved ground state. Within the pair, the beams have opposite polarizations, with their frequencies separated by less than the MOT transition linewidth. This approach is in contrast with previously realized blue molecular MOT schemes, which use at most one frequency component per ground state.
Using our 1+2 MOT scheme, we realize a peak molecular density for CaOH of $8(2)\times10^8~\text{cm}^{-3}$, with $\sim$$2{,}900$ molecules trapped at a cloud radius of $59(5)~\text{\textmu m}$.
This density is more than two orders of magnitude larger than we achieve with a red MOT for CaOH~\cite{CaOH_MOT} and is the highest density reported for a molecular MOT to date.
The temperature of the 1+2 MOT is~$170(10)~\text{\textmu K}$, slightly above the Doppler temperature for CaOH of~$140~\text{\textmu K}$. 
To compare the 1+2 MOT scheme with existing blue MOT schemes, we implement a ``1+1'' MOT scheme for CaOH, using the standard single frequency component for each of the two resolved ground states. Significantly weaker compression is found for the 1+1 scheme, realizing a density of $5(1) \times 10^7~\text{cm}^{-3}$.
The use of a 1+2 scheme was motivated by our studies of molecular MOTs using Monte Carlo simulations based on the stochastic Schr\"{o}dinger equation~\cite{molmer1993monte}. We find good agreement between our experimental data and the simulation results.

\begin{figure}[t!]
    \centering
    \includegraphics{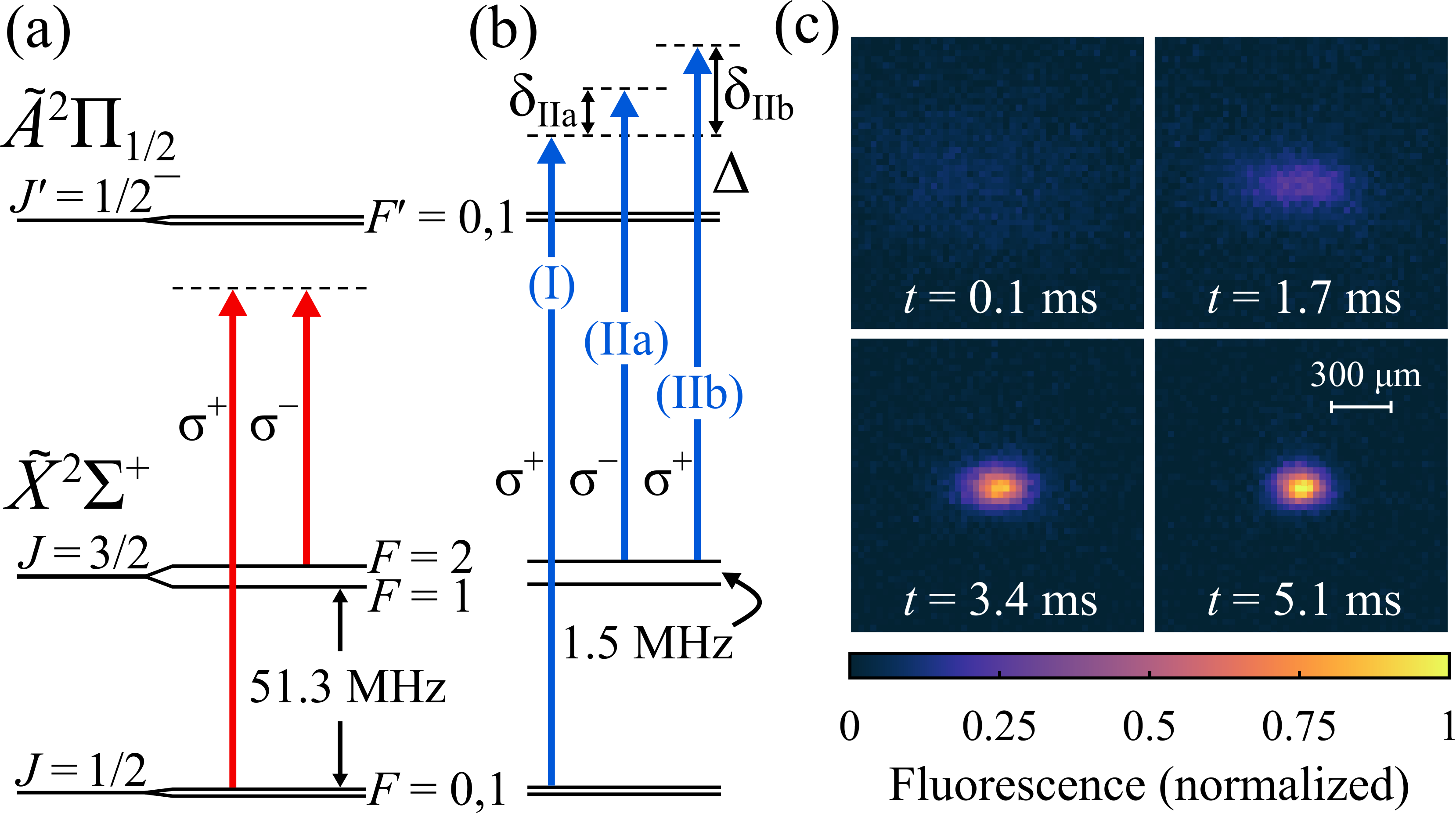}
    \caption{(a) Level structure and laser configuration for the CaOH red (RF) MOT. (b) CaOH blue (DC) MOT laser configuration. The single-photon detuning is denoted by $\Delta$, and the two-photon detunings for the two MOT beams addressing the $\ket{J=3/2}$ level are denoted by $\delta_\text{IIa}$, and $\delta_\text{IIb}$. (c) Images of the blue-detuned CaOH MOT taken with $2~\text{ms}$ in-situ images following a compression time of $t$.}
    \label{fig:figure1}
\end{figure}

In our experiment, CaOH molecules produced from a cryogenic buffer gas beam source are first radiatively slowed and then captured in a red-detuned radio-frequency (RF) MOT, which is created using MOT beams with frequencies less than the frequency of the $\tilde{A}^2 \Pi_{1/2}(J'=1/2) \leftarrow \tilde{X}^2\Sigma^+(N=1)$ electronic transition.
(Since blue-detuned light gives rise to Doppler heating around the MOT capture velocity, blue MOTs do not provide initial capture and cooling.)
The CaOH level structure and red MOT laser configuration are shown in Fig.~\ref{fig:figure1}(a).
The $\ket{N=1}$ state contains two spin-rotation states, $\ket{J=1/2}$ and $\ket{J=3/2}$, split by $\sim$$51~\text{MHz}$. Each spin-rotation state is further split into a pair of hyperfine states spaced by $\sim$$10~\text{kHz}$ in $\ket{J=1/2}$ and $\sim$$1.5~\text{MHz}$ in $\ket{J=3/2}$. This splitting is unresolved compared to the linewidth of the MOT transition of $\Gamma = 2\pi \times 6.4~\text{MHz}$, and the red MOT light contains single frequency components to address each spin-rotation component (Fig.~\ref{fig:figure1}(a)). The slowing light used to bring CaOH molecules to the capture velocity of the red MOT contains additional laser frequencies that repump molecules from twelve excited vibrational states of $\tilde{X}^2\Sigma^+$ back into the cycling transition, resulting in a ``photon budget'' of $\sim$$13{,}500$ scattered photons per $1/e$ loss of the molecular population to unaddressed vibrational levels.
These repumping lasers pass through the MOT region and remain active during the full experimental sequence. Further details of the red MOT, which is operating as an RF MOT, and its laser configuration and operation are in ref.~\cite{CaOH_MOT}. After the initial capture of molecules, the MOT magnetic field gradient is ramped from 8.5 G/cm to 23.6 G/cm. Following a $10~\text{ms}$ wait time where spatial compression takes place, the Gaussian rms width $\sigma$ of the red MOT is measured to be $0.53(3)~\text{mm}$ and the peak molecular density is $2.0(3) \times 10^{6}~\text{cm}^{-3}$.

We realize a blue MOT by loading CaOH from a red MOT. First, the red MOT is switched off by turning off both the MOT light and the RF magnetic field in $\sim$$130~\text{\textmu s}$. 
The MOT light beams are then turned back on in a ``$\Lambda$-cooling'' configuration to perform $1~\text{ms}$ of $\Lambda$-enhanced gray molasses cooling in free space, resulting in a molecular temperature of $\sim$$34~\text{\textmu K}$.
Details on $\Lambda$-cooling for CaOH are in ref.~\cite{CaOH_ODT}. This $\Lambda$-cooling step prevents the molecules from leaving the MOT region during the time when the blue MOT is turned on and ensures that the molecules are within the relatively lower capture velocity of the blue MOT. The blue MOT is then created by switching the MOT laser configuration to the one shown in Fig.~\ref{fig:figure1}(b) and linearly ramping up a DC magnetic field in $4~\text{ms}$. 
We observe negligible molecule loss during transfer to the blue MOT.
In the blue MOT laser configuration, a single frequency component (I), blue-detuned from the MOT transition by $\Delta$, addresses the $\ket{J=1/2}$ state while two closely-spaced frequency components of opposite polarization, denoted by (IIa) and (IIb), address the $\ket{J=3/2}$ state. Referenced to the $F=2$ hyperfine state, the detunings of (IIa) and (IIb) are $\Delta + \delta_\text{IIa}$ and $\Delta + \delta_\text{IIb}$, respectively, where $\delta_\text{IIa}$ and $\delta_\text{IIb}$ are the two-photon detunings for the (IIa) and (IIb) components with respect to (I). This configuration with both (I) and the (IIa)-(IIb) pair is referred to as the 1+2 scheme.

Fig.~\ref{fig:figure1}(c) shows in-situ images taken throughout the compression of the molecular cloud while the 1+2 MOT light is applied. The ramped (axial) magnetic field gradient is $B'_z = 78~\text{G/cm}$.
The frequency parameters for the MOT light are $\Delta = 7.4~\text{MHz}$, $\delta_\text{IIa} = -1.0~\text{MHz}$, $\delta_\text{IIb} = 0.75~\text{MHz}$ and were chosen to maximize the compression of the molecular cloud. We note that, since the $\ket{J=3/2, F=1}$ is only $\sim$$1.5~\text{MHz}$ lower in energy compared to the $\ket{J=3/2, F=2}$ state, the two-photon detunings with respect to this state are of a similar scale, though negative for both (IIa) and (IIb). The total 1+2 MOT light intensity is ramped linearly from $I = 23.5~\text{mW/cm}^2$ to $I = 18.2~\text{mW/cm}^2$ during the magnetic field ramp. The fractions of the total intensity among the three frequency components are $37\%$, $28\%$, and $35\%$ for (I), (IIa), and (IIb), respectively.
The 1+2 MOT is fully compressed after $5~\text{ms}$ and reaches a final size of $\sigma = 59(5)~\text{\textmu m}$.
Using the time-of-flight expansion method, the molecular temperature is measured to be $T = 170(10)~\text{\textmu K}$. Details on the size and temperature measurement methods are given in the Supplemental Material~\cite{supplement}. After the 1+2 MOT compression, $\sim$$60\%$ of the molecules that were in the red MOT remain, corresponding to $\sim$$2{,}900$ molecules. With the measured final MOT size, this results in a peak number density of $n_0 = 8(2) \times 10^8~\text{cm}^{-3}$.
The molecule loss is almost entirely from branching losses to vibrational dark states during photon cycling due to the finite photon budget. 
Given our scattering rate of $0.90(3)~\text{MHz}$ and a photon budget of $13{,}500$ photons, this loss can be estimated theoretically as $e^{-0.9~\text{MHz} \, \times \, 5~\text{ms} / 13{,}500} = 71\%$.

We quantify the cooling and trapping forces in the MOT by modelling the MOT as a damped harmonic oscillator with force $F = -\alpha v - k r$. Here $r$ is position, $v$ is velocity, $\alpha$ is the damping coefficient, and $k$ is the spring constant.
Using the equipartition theorem, the spring constant can be extracted from the size and temperature according to $k = k_B T/\sigma^2$~\cite{Rb}.
We find $k = 6(1)\times 10^{-19}~\text{N/m}$ for the 1+2 MOT, which is more than an order of magnitude larger than spring constants achieved in previously demonstrated blue-detuned MOTs~\cite{YO, jorapur2023, CaF}.

\begin{figure*}[t!]
    \centering
    \includegraphics{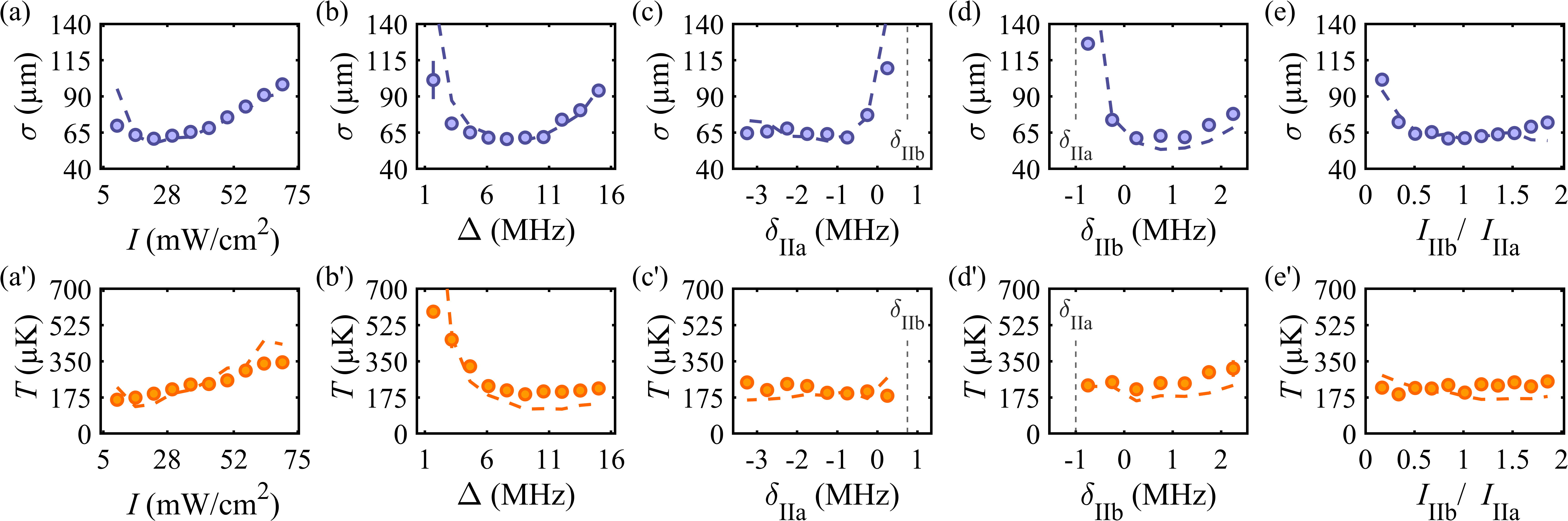}
    \caption{1+2 MOT (a-e) width $\sigma$ and (a'-e') temperature $T$ following $5~\text{ms}$ compression,
    as a function of (a,a') total power $I$; (b,b') one-photon detuning $\Delta$; (c,c') two-photon detuning of the (IIa) component $\delta_\text{IIa}$; (d,d') two-photon detuning of the (IIb) component $\delta_\text{IIb}$; (e,e') intensity ratio between (IIb) and (IIa) (with total $I$ fixed). The fixed parameters for each scan are $I = 18.2~\text{mW/cm}^2$, $\Delta = 7.4~\text{MHz}$, $\delta_\text{IIa} = -1~\text{MHz}$, $\delta_\text{IIb} = 0.75~\text{MHz}$, and $I_\text{IIb} / I_\text{IIa} = 1.25$. The MOT magnetic field gradient is $78~\text{G/cm}$. Dashed curves are simulation results.
    }
    \label{fig:figure3}
\end{figure*}

\begin{figure}[b!]
    \centering
    \includegraphics{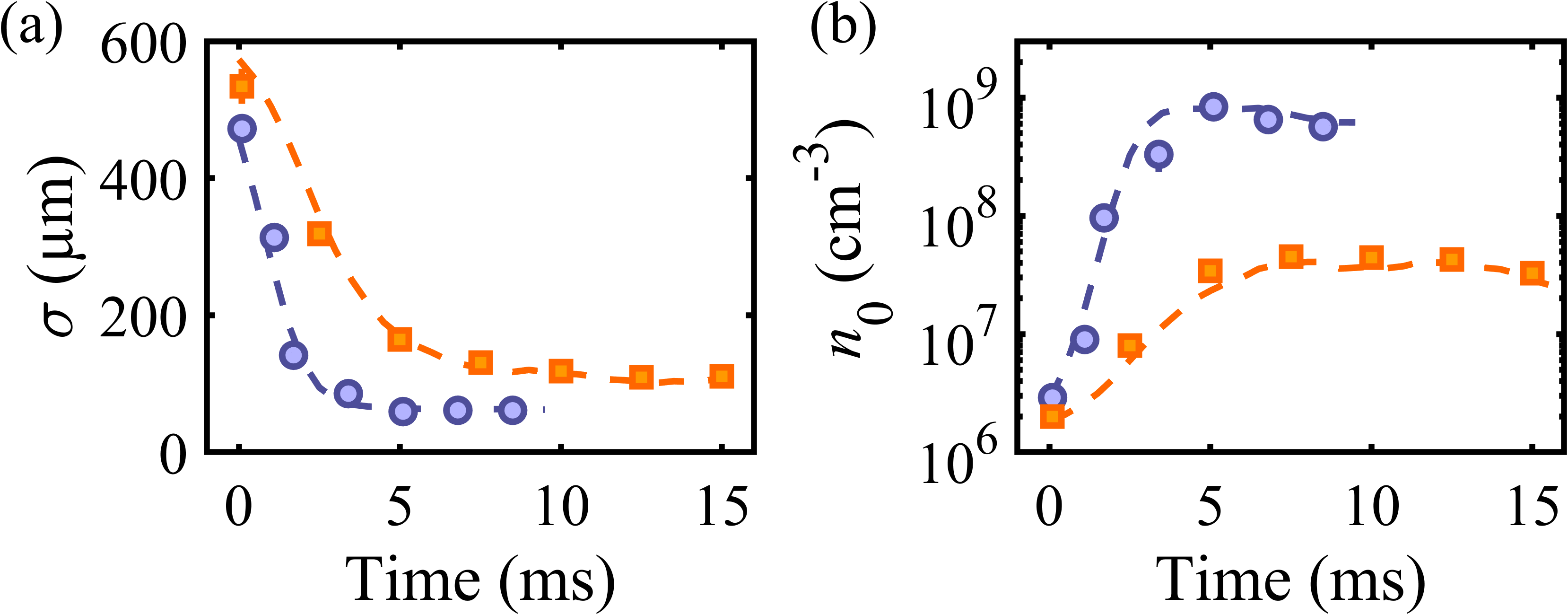}
    \caption{(a) MOT width $\sigma$ and (b) MOT peak number density $n_0$ as a function of compression time for the 1+2 MOT (blue circles) and the 1+1 (A) MOT (orange squares). Dashed curves are simulation results.}
    \label{fig:figure2}
\end{figure}

We further characterize the 1+2 MOT by measuring the MOT width $\sigma$ and temperature $T$ as a function of several MOT parameters (Fig.~\ref{fig:figure3}). 
The intensity and single-photon detuning ($\Delta$) data show clear minima in the MOT size~(Fig.~\ref{fig:figure3}(a,b)), while the temperature is minimized at the lowest light intensity and for detunings $\Delta \gtrsim 7~\text{MHz}$~(Fig.~\ref{fig:figure3}(f,g)). 
We find that the 1+2 MOT size has a strong dependence on $\delta_\text{IIa}$ and $\delta_\text{IIb}$ (Fig.~\ref{fig:figure2}(c,d)), with the smallest $\sigma$ achieved for $\delta_\text{IIa} < 0$ and $0 < \delta_\text{IIb} < 1.5~\text{MHz}$. 
The most obvious effect, however, is that when $\delta_\text{IIa}$ and $\delta_\text{IIb}$ are tuned close to each other, $\sigma$ increases dramatically, i.e., the trapping force vanishes.
The variation of the relative power ratio between (IIb) and (IIa) demonstrates that, as the intensity of (IIb) is increased from zero, the cloud size shrinks to the minimum size~(Fig.~\ref{fig:figure3}(e)). The temperature is relatively insensitive to both $\delta_\text{IIa}$ and $\delta_\text{IIb}$ as well as the power ratio between (IIa) and (IIb)~(Fig.~\ref{fig:figure3}(h,i,j)).

The experimentally observed variation of the 1+2 MOT size and temperature are compared to numerical predictions from Monte-Carlo simulations, shown in Fig.~\ref{fig:figure3} as dashed curves. Briefly, these simulations are performed by initializing CaOH molecules from position and velocity distributions coinciding with the $\Lambda$-cooled molecule cloud (following the red MOT), and then numerically solving the stochastic Schr\"{o}dinger equation~\cite{molmer1993monte} for each molecule in the presence of the blue MOT light and fields. 
Simulated MOT sizes and temperatures are derived as a function of time from the solved trajectories. The simulation results show excellent agreement with the experimental data. Additional details on the simulations are provided in the Supplemental Material~\cite{supplement}.

We also create a blue MOT that employs laser frequency (I) and only one of either (IIa) or (IIb). 
We refer to these 1+1 schemes as (A) using (I) and (IIa), and (B) using (I) and (IIb).
A difference between (A) and (B) is that the polarization for the component addressing the $\ket{J=3/2}$ state is reversed.
The 1+1 (A) MOT is the approach used in the blue MOTs created for YO and CaF and follows the conventional MOT configuration where the frequency separation between each of the components of the MOT light is significantly greater than the MOT transition linewidth~\cite{YO, CaF}.
We find significantly reduced spatial compression in both of the 1+1 schemes.
In Fig.~\ref{fig:figure2}, we compare the compressions of the 1+2 and 1+1 (A) MOTs.
The compression predicted by the numerical simulations for the MOTs are shown as dashed curves and are in excellent agreement with the experimentally observed sizes and compression timescales.
At optimal parameters for compression, the 1+1 (A) MOT achieves a final size of $\sigma = 110(4)~\text{\textmu m}$ after 15~\text{ms} of compression, which is $\sim$$2\times$ larger than the size achieved with the 1+2 MOT. As with the 1+2 MOT, the gradient is ramped to $B_z' = 78~\text{G/cm}$ over the first 4~\text{ms}. The temperature of the 1+1 (A) MOT is $T = 160(10)~\text{\textmu K}$, which is similar to the temperature of the 1+2 MOT. For the 1+1 (A) MOT, $k = 1.8(1) \times 10^{-19}~\text{N/m}$, about $\sim$$3.5\times$ smaller than the spring constant for the 1+2 MOT.
The scattering rate of the 1+1 (A) MOT is $0.7(1)~\text{MHz}$, only $\sim$$20\%$ smaller than that of the 1+2 MOT, so the large difference in spring constant implies that the 1+2 MOT achieves a significantly larger restoring force per photon scattered.
The peak molecular density of the 1+1 (A) MOT is $n_0 = 5(1) \times 10^7~\text{cm}^{-3}$, a factor of $\sim$$16$ smaller compared to the 1+2 MOT.

We note that, for the 1+1 schemes, a strong dependence of the spatial compression of the MOT cloud on $\delta$ was observed (here $\delta$ refers to $\delta_\text{IIa}$ and $\delta_\text{IIb}$ for each of the respective 1+1 (A) and (B) schemes.)
We find for the 1+1 (A) MOT that the sign of the MOT force is inverted around $\delta \approx 0$, transitioning from a trapping to an anti-trapping force~\cite{supplement}. 
The strong dependence of the trapping force on $\delta$, not only for the magnitude but also the sign of the force, suggests that two-photon effects may play an important role for the restoring forces achieved, perhaps in both the 1+1 and 1+2 MOTs.
A more general investigation of the effects of two-photon resonances on MOT forces in type-two systems is warranted.

\begin{figure}[t!]
    \centering
    \includegraphics{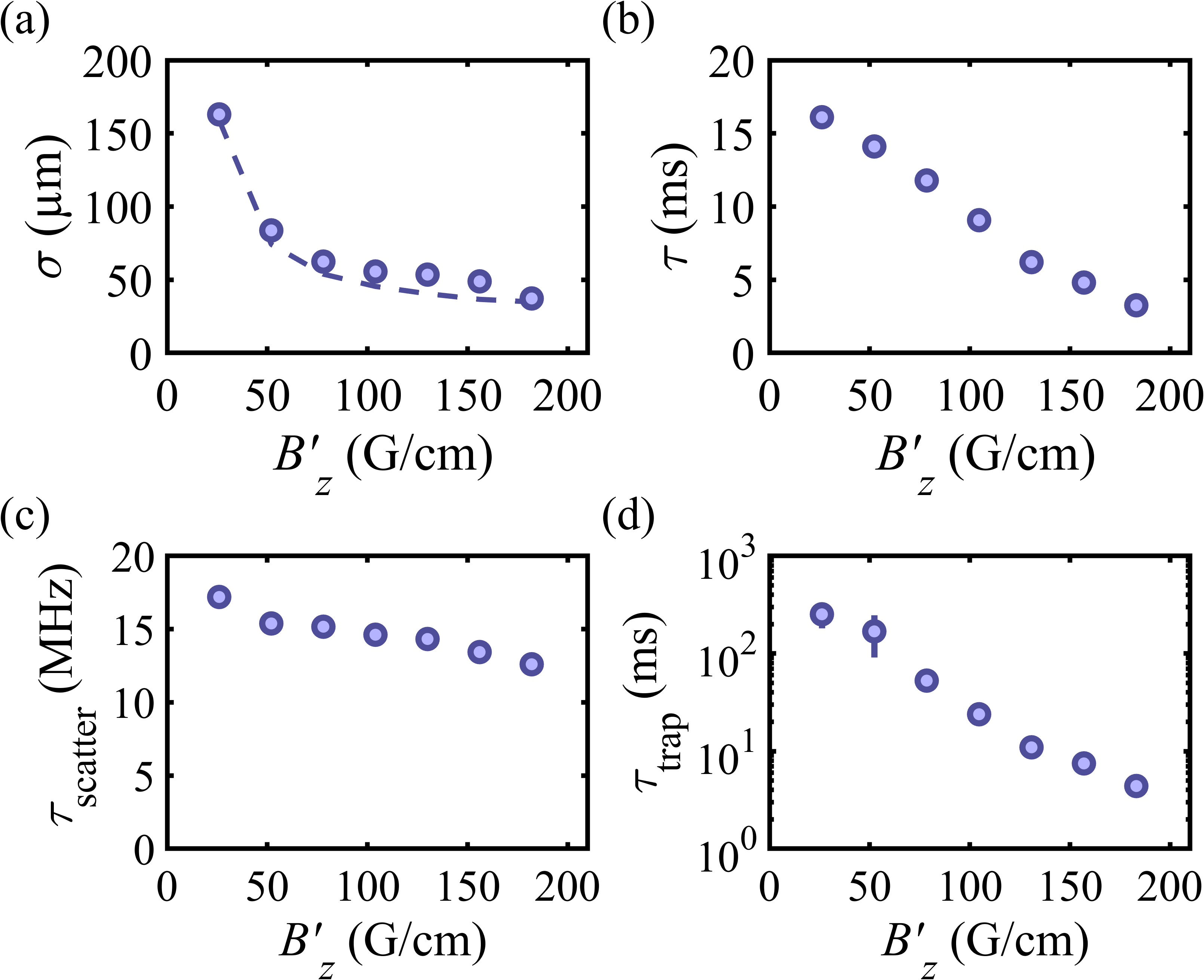}
    \caption{1+2 MOT (a) width $\sigma$ (dashed line shows simulation results), (b) lifetime $\tau$, (c) lifetime $\tau_\text{scatter}$ from branching loss to vibrational dark states, and (d) lifetime $\tau_\text{trap}$ from trap losses only, as a function of magnetic field gradient.}
    \label{fig:figure5}
\end{figure}

We characterize the performance of the 1+2 MOT versus magnetic field gradient $B'_z$. As shown in Fig.~\ref{fig:figure5}(a), the compression continues to increase with gradient and, with $B'_z = 182~\text{G/cm}$, the cloud is compressed to a size of $37(2)~\text{\textmu m}$. The temperature remains roughly constant at $\sim$$150$--$200~\text{\textmu K}$ for the gradients explored. Fig.~\ref{fig:figure5}(b) shows the lifetime $\tau$ of the molecular cloud at different gradients, measured by holding the compressed molecule cloud at each gradient for a varying duration and observing the number of remaining molecules. With a gradient of $B'_z = 78~\text{G/cm}$, we measure $\tau = 11.8(2)~\text{ms}$. 
Due to our current (technologically limited) photon budget of $\sim$$13{,}500$ photons, $\tau$ arises both from loss due to branching into dark vibrational states ($\tau_\text{scatter}$) and loss from molecules leaving the trap ($\tau_\text{trap}$).
The total lifetime is given by $\tau = (1/\tau_\text{scatter} + 1/\tau_\text{trap})^{-1}$. To isolate $\tau_\text{scatter}$ and $\tau_\text{trap}$, we first measure the photon scattering rate $R_\text{sc}$ at each gradient by repeating the same lifetime measurement but with a photon budget limited to $\sim$$1400$ photons such that $\tau \approx \tau_\text{scatter} \ll \tau_\text{trap}$. We calculate $\tau_\text{scatter} = 13{,}500 / R_\text{sc}$ as a function of gradient (Fig.~\ref{fig:figure5}(c)). Using $\tau$ and $\tau_\text{scatter}$, we also calculate $\tau_\text{trap} = (1/\tau - 1/\tau_\text{scatter})^{-1}$ as a function of gradient (Fig.~\ref{fig:figure5}(d)). With $B'_z = 78~\text{G/cm}$, we find $\tau_\text{trap} = 53(7)~\text{ms}$, implying that, at this gradient, the lifetime is limited by $\tau_\text{scatter}$. As the gradient is increased, the contribution from $\tau_\text{trap}$ becomes increasingly important, and with $B'_z \approx 125~\text{G/cm}$, $\tau$ has roughly equal contributions from both $\tau_\text{scatter}$ and $\tau_\text{trap}$. We note that the trap-limited lifetime $\tau_\text{trap}$ in the 1+1 (A) MOT is significantly shorter than that of the 1+2 MOT. For the 1+1 (A) scheme, with $B'_z = 78~\text{G/cm}$, we measure $\tau = 7(1)~\text{ms}$ and $\tau_\text{scatter} = 20(6)~\text{ms}$, which results in $\tau_\text{trap} = 12(3)~\text{ms}$.

The mechanism underlying the improved compression of the 1+2 MOT is not yet fully understood. However, our simulations suggest a possible explanation. When both of the (IIa) and (IIb) components are present, the MOT light addressing the $\ket{J=3/2}$ state can, in each MOT axis, be decomposed into two slowly moving standing waves (``walking waves'') with opposite circular polarizations and directions of motion.
Due to spatially-dependent Zeeman shifts in the MOT, a molecule may preferentially interact with one of the opposing walking waves depending on its location.
In a reference frame comoving with a walking wave, the molecule experiences Sisyphus-like cooling owing to the intensity gradient in the standing wave~\cite{devlin2016three}. In the lab frame, the cooling force becomes a pushing force. 
By an appropriate choice of laser polarizations and magnetic field direction, this mechanism could provide a spatially dependent restoring force.
We provide a detailed explanation of this hypothesized trapping mechanism in the Supplemental Material~\cite{supplement} for a toy model with a single type-two transition, $\ket{F=1} \rightarrow \ket{F'=0}$, addressed by $\sigma^+$ and $\sigma^-$ laser components that are close in frequency, as in the 1+2 scheme. Numerical simulations of molecule trajectories show that, in this toy model, molecules concentrate around the speed of the walking waves during compression, indicating a cooling force in a moving frame.
Further investigation is required to explore if the proposed mechanism takes place in the CaOH MOT and explains our results.
We reserve this to future work.

It is interesting to explore whether 1+2 type schemes can lead to improved spatial compression in type-two MOTs for other molecules. 
Using our simulations, we develop promising MOT schemes similar to our 1+2 MOT for CaF and YO. Details of these simulations are provided in the Supplemental Material~\cite{supplement}. 

In summary, we have demonstrated a blue-detuned MOT for a polyatomic molecule, CaOH. We have devised and studied a ``1+2'' MOT light scheme for CaOH that uses two closely-spaced frequency components to address the same ground state and achieves significantly improved spatial compression compared to previously demonstrated blue-detuned molecular MOTs.
Our results also suggest that two-photon effects can play an important role for MOT forces in type-two systems and that these effects should be taken into account when developing schemes for molecular MOTs. Based on numerical Monte-Carlo simulations, similar 1+2 MOT schemes should be applicable to other molecules such as CaF, YO, SrF, and SrOH, and possibly larger species. The spatial compression and concomitant increase in molecular number density offered by this blue-detuned MOT should improve the loading of optical traps. With better loading into optical dipole traps, one can expect that collisions of CaOH molecules in the bulk can be studied. Approaches have been suggested for realizing collisional shielding in CaOH, which could be implemented to achieve evaporative cooling~\cite{augustovivcova2019ultracold}. Higher densities should also enable more robust loading of individual tweezer traps, which could aid in scaling to larger molecular array sizes.

We acknowledge Yicheng Bao and Jiaqi You for valuable discussions in developing the numerical Monte-Carlo simulations. This work was supported by the AFOSR, NSF, ARO, DOE Quantum Systems Accelerator (QSA), and the CUA (PHY-2317134). GKL and LA acknowledge support from the HQI, and PR acknowledges support from the NSF GRFP.

\bibliography{CaOH_BlueMOT_References}

\end{document}

\beginsupplement

\title{Supplemental Material}

\maketitle

\section{Size and temperature measurements}
In the experiment, the size $\sigma$ of the blue MOT is determined by capturing a brief in-situ image (2~ms) at the end of the blue MOT sequence, and fitting the image data to a 2D Gaussian function. The temperature is determined using a time-of-flight (TOF) sequence. Following the compression in the blue MOT, the molecule cloud is allowed to freely expand for various durations, after which it is imaged using $\Lambda$-cooling light. The fitted Gaussian rms width of the expanded cloud at different TOF time $t$ is used to determine the temperature of the cloud, using the relation
\begin{equation}
    \sigma(t) = \sqrt{\sigma_0^2 + \frac{k_B T}{m} t^2},
\end{equation}
where $m$ is the mass of CaOH and $\sigma_0$ is the cloud size at $t = 0$. 

\section{Numerical Simulations}
\begin{figure}
    \centering
    \includegraphics[width=1\linewidth]{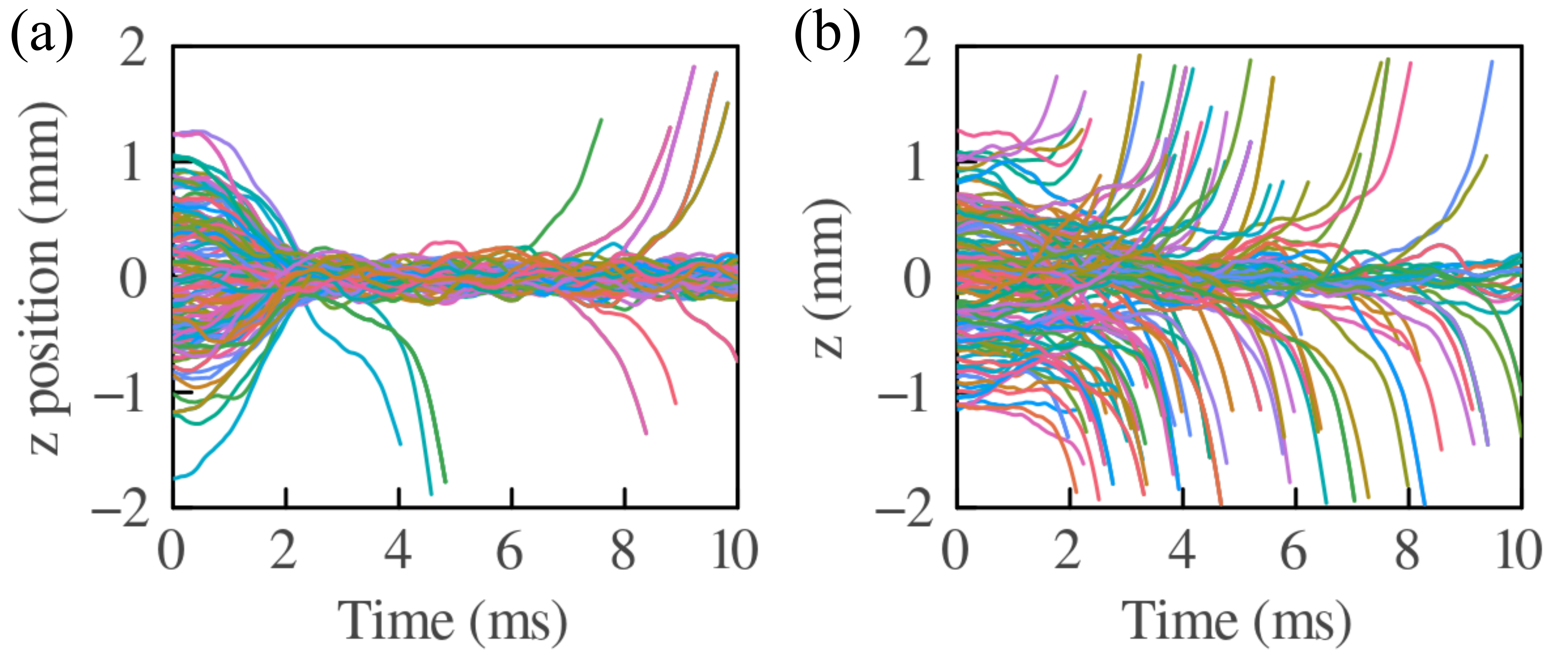}
    \caption{Position along the $\hat{z}$ direction for simulated particle trajectories for the CaOH (a) 1+2 MOT and the (b) 1+1 MOT.}
    \label{fig:trajectories}
\end{figure}

We use Monte Carlo simulations to model the blue MOT experiment and achieve good agreement with our experimental observations. Dynamics of the internal molecular state evolution is modeled based on the stochastic wavefunction approach~\cite{Dalibard_stochastic}. This involves propagating the wavefunction of the 16 photon-cycling states of the CaOH MOT transition according to the Schr\"{o}dinger equation while incorporating spontaneous emissions as stochastic events. 

The average dipole force is computed from the expectation value of the force operator
\begin{equation}
    \mathbf{f} = \langle \hat{\mathbf{f}} \rangle = \langle \mathbf{\nabla} \hat{H} \rangle,
\end{equation}
which we use to update the positions and velocities of the molecules at each time step during the evolution. Here $\hat{H}$ is the molecule-light interaction Hamiltonian. A momentum kick of magnitude $\hbar k$ is applied in a random direction every time spontaneous emission occurs, and momentum diffusion due to fluctutations of the dipole force operator is added per ref.~\cite{dalibard1985atomic}. Further details regarding the implementation will be discussed in a future work.

For the simulations presented in this paper, we initialize the molecules randomly within a cloud of $\sim$$600~\text{\textmu m}$ radius and at a temperature of $\sim$$35~\text{\textmu K}$. Then the same blue MOT sequence used in the experiment is applied. Specifically, for the optimal 1+2 blue MOT configuration, this involves ramping up the magnetic field gradient from $0$ to $78~\text{G/cm}$ over a duration of 4 ms, while simultaneously reducing the intensities from $I = 23.5~\text{mW/cm}^2$ to $I = 18.2~\text{mW/cm}^2$. Simulated particle trajectories for the 1+2 and 1+1 MOTs are illustrated in Fig.~\ref{fig:trajectories}. The final size and temperature of the molecule cloud are determined by fitting the position and velocity distributions along each direction to a Gaussian function.

The simulations use the laser frequencies and relative intensity ratios taken directly from the experiment, as reported in the main text. The actual, total intensity seen by the molecules can, however, be significantly smaller than the measured peak intensity of the MOT beams. This is due to the fact that the MOT beams have a Gaussian profile with a $1/e^2$ diameter of $8~\text{mm}$ and the location of the compressed molecule cloud with respect to the MOT beams is not fully known. We find that by setting the intensities used in the simulation to $\sim$$1/2$ of the measured peak intensity, the scattering rates predicted by simulations match the experimentally measured scattering rates for both the 1+2 and 1+1 MOT schemes. This scaling factor is applied to all simulations (with the exception of the YO and CaF simulations described in sec. IV).

\section{Two-photon detuning for 1+1 schemes}
\begin{figure}
    \centering
    \includegraphics{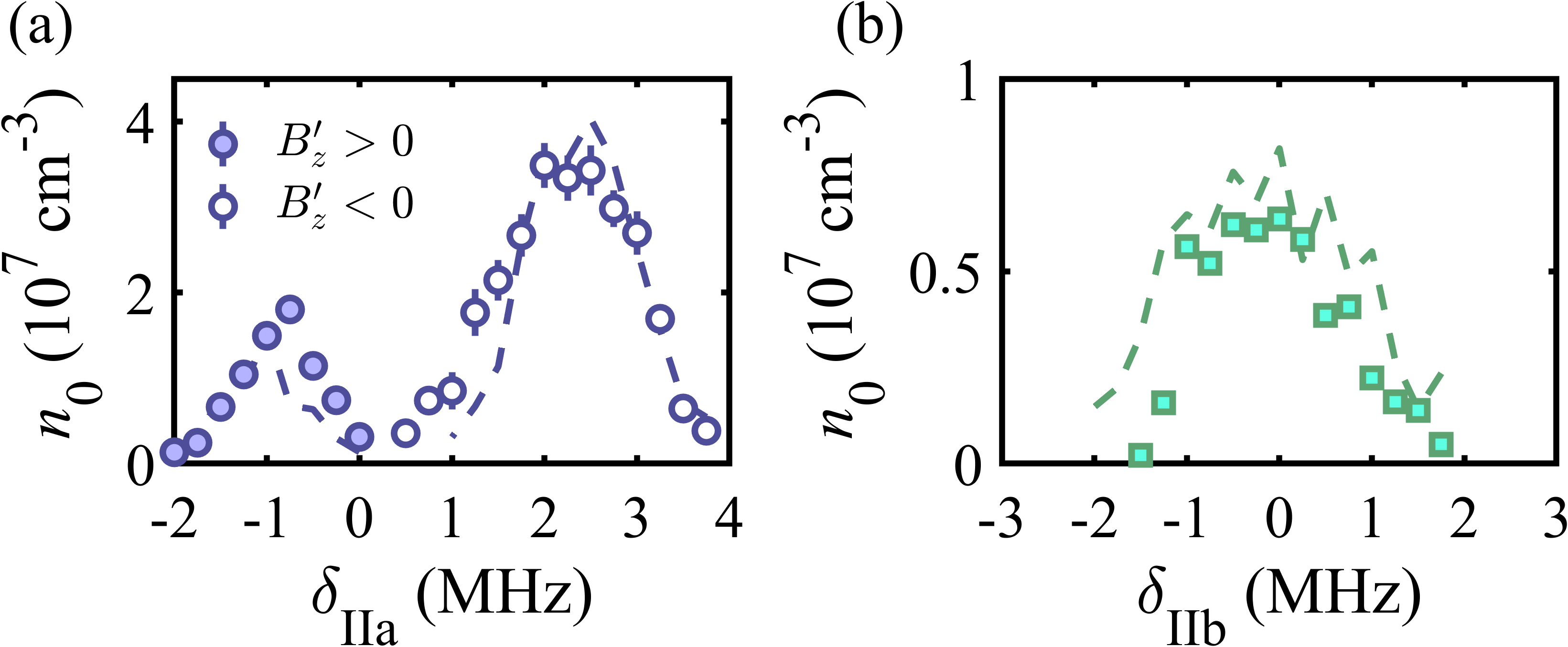}
    \caption{(a) MOT peak number density $n_0$ versus two-photon detuning $\delta_\text{IIa}$ for the 1+1 (A) MOT, for both orientations of magnetic field gradient. (b) MOT peak number density $n_0$ versus two-photon detuning $\delta_\text{IIb}$ for the 1+1 (B) MOT. Dashed curves are simulation results.}
    \label{fig:figure1s}
\end{figure}

The MOT density as a function of $\delta_\text{IIa}$ and $\delta_\text{IIb}$ for the 1+1 (A) and 1+1 (B) schemes is shown in Fig.~\ref{fig:figure1s}. For (A), we find that a trapping force is achieved only around a narrow range of two-photon detunings $\delta_\text{IIa} = -2$ to $0~\text{MHz}$ and that the trapping converts into anti-trapping around $\delta_\text{IIa} \approx 0~\text{MHz}$. The anti-trapping is evident from the fact that, for $\delta_\text{IIa} > 0~\text{MHz}$, a MOT can be achieved by reversing the direction of the magnetic field gradient. For (B), a trapping force is observed in the range $\delta_\text{IIb} = -2$ to $2~\text{MHz}$, but no anti-trapping is observed. For both cases, the molecular density is much less sensitive to the single-photon detuning $\Delta$ and follows a similar trend to that of the 1+2 scheme, as shown in Fig.~3(b) in the main text.
The observed dependence on two-photon detuning is reproduced for both (A) and (B) in the numerical simulations, as indicated by the dashed curves. We give a brief hypothesis for the observed results at the end of sec. V.

\section{Simulations for YO and CaF}

Here we provide details on numerical simulations of blue MOTs for CaF and YO. We simulate the MOT schemes from refs.~\cite{CaF, YO} realized for CaF and YO, which, like CaOH, uses a $\tilde{X}^2\Sigma^+(N=1) \rightarrow \tilde{A}^2 \Pi_{1/2}(J'=1/2)$ transition.

\begin{figure}
    \centering
    \includegraphics{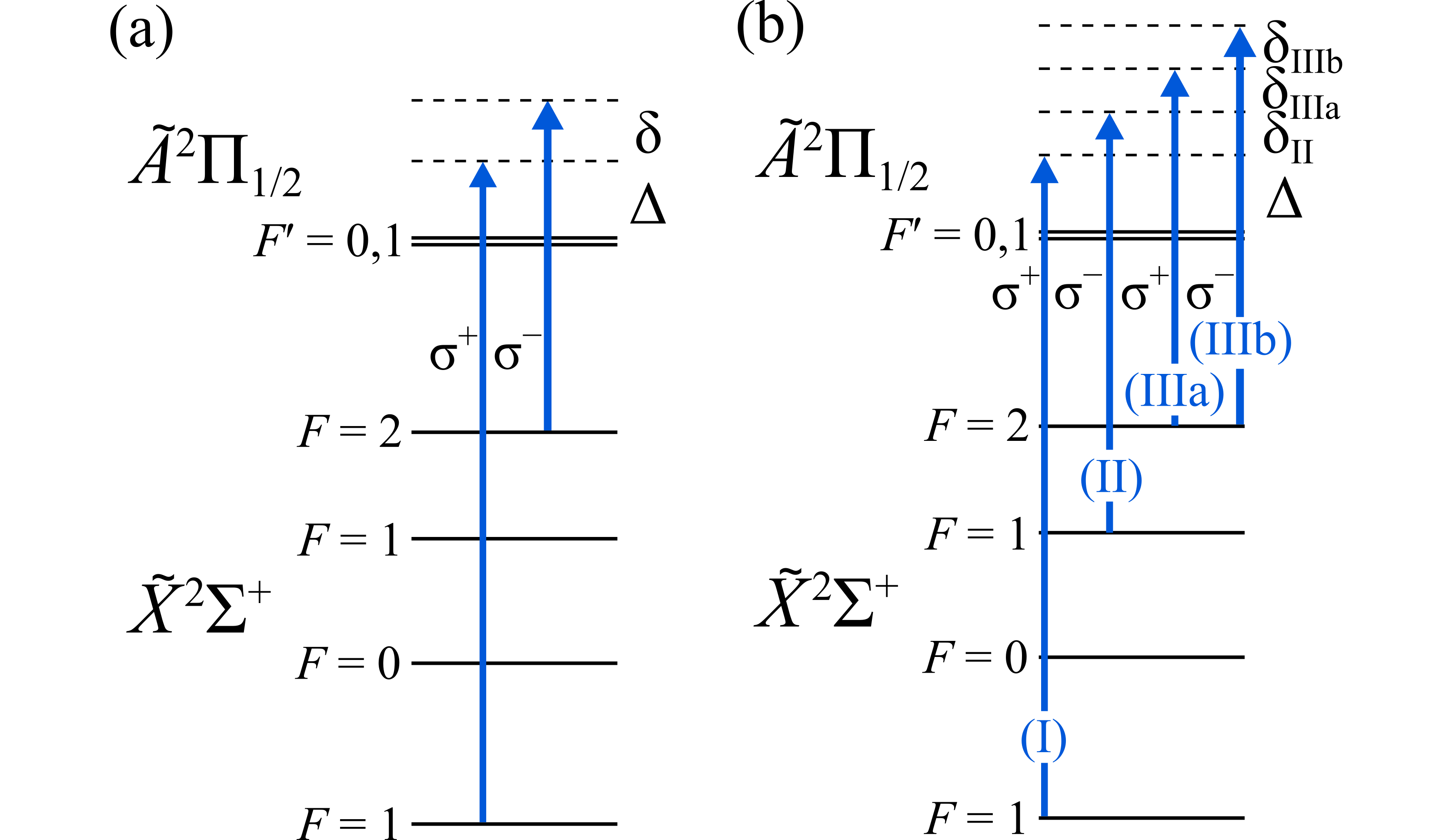}
    \caption{(a) Blue MOT scheme for CaF used in ref.~\cite{CaF}. (b) Proposed blue MOT scheme for CaF using a similar scheme to the 1+2 blue CaOH MOT.}
    \label{fig:figure3s}
\end{figure}

\begin{figure}
    \centering
    \includegraphics{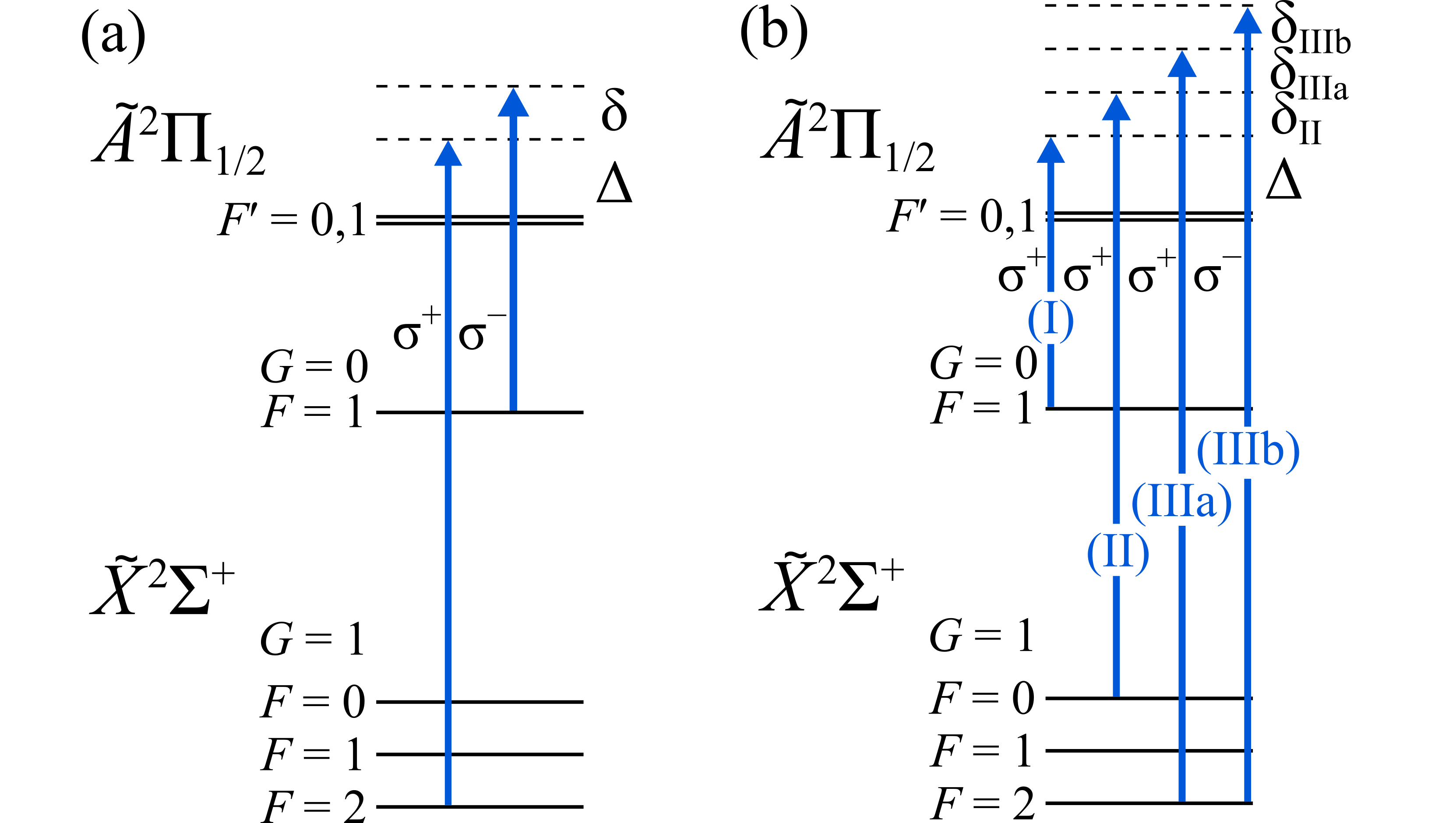}
    \caption{
    (a) Blue MOT scheme for YO used in ref.~\cite{YO}. (b) Proposed blue MOT scheme for YO using a similar scheme to the 1+2 blue CaOH MOT.}
    \label{fig:figure2s}
\end{figure}

Fig.~\ref{fig:figure3s} shows the scheme used to realize a blue MOT for CaF in ref.~\cite{CaF}, as well as a proposed scheme for CaF similar to the 1+2 scheme for CaOH. In our simulation of the previously realized scheme (Fig.~\ref{fig:figure3s}(a)), we use a total intensity of $I = 5.8~\text{mW/cm}^2$ (split $1$-to-$1.4$ between the two components addressing $\ket{F=1}$ and $\ket{F=2}$), $\Delta = 23.8~\text{MHz}$, $\delta = -0.75~\text{MHz}$, and $B_z' = 14.6~\text{G/cm}$. This scheme is similar to our 1+1 (A) scheme for CaOH.
Our simulations for this scheme predict a MOT size of $\sigma_\text{sim} \approx 230~\text{\textmu m}$ and a temperature of $T_\text{sim} \approx 100~\text{\textmu K}$, which is in reasonable agreement with experimental results from ref.~\cite{CaF} for similar parameters. 
The proposed scheme (Fig.~\ref{fig:figure3s}(b)) has two closely-spaced frequency components (IIIa) and (IIIb) that address the $\ket{F=2}$ state in CaF. For this scheme, the simulations give a predicted size of $\sim$$45~\text{\textmu m}$ and a temperature of $\sim$$100~\text{\textmu K}$. We find that in order to achieve this high compression, an additional frequency component (II) added to the higher $\ket{F=1}$ state is also required. This may be to avoid gray-molasses type heating from the (III) components, which are red-detuned from the $\ket{F=1}$ state. Notably, we still only find that high compression can be achieved using both of the (III) components.
The specific parameters for the 1+2 CaF scheme are $I = 29~\text{mW/cm}^2$ (split $22\%/58\%/10\%/10\%$ among the (I)/(II)/(IIIa)/(IIIb) components), $\Delta = 9.0~\text{MHz}$, $\delta_\text{II} = 3.0~\text{MHz}$, $\delta_\text{IIIa} = -0.6~\text{MHz}$, $\delta_\text{IIIb} = 0.6~\text{MHz}$, and $B = 78~\text{G/cm}$. 

Fig.~\ref{fig:figure2s}(a) shows the MOT scheme used in ref.~\cite{YO} to create a blue MOT for YO. We note that, for these YO simulations, we found that our approach for estimating momentum diffusion due to dipole force fluctuations resulted in temperatures significantly higher than the temperatures observed in ref.~\cite{YO}. Indeed, when $\Lambda$-cooling is promiment and molecules spend significant time in velocity-dependent dark states (as might be expected in the low-gradient YO MOT), our momentum diffusion estimation is possibly no longer appropriate. For this simulation we  therefore instead use a fixed momentum diffusion constant to roughly match the temperature observed in the experimental data of ref.~\cite{YO}. With this approach, our simulations for the previous YO scheme predict a size of $\sigma_\text{sim} \approx 180~\text{\textmu m}$ and a temperature of $T_\text{sim} \approx 80~\text{\textmu K}$, which are in reasonable agreement with the experimentally measured values given in ref.~\cite{YO}. Fig.~\ref{fig:figure2s}(b) shows a proposed scheme for a YO MOT similar to the proposed CaF scheme as well as the 1+2 CaOH MOT configuration described in the main text. In this scheme for YO, the $\ket{F=2}$ state is addressed by two closely-spaced frequency components (IIIa) and (IIIb) of opposite polarization. An additional component (II) is added to address the $F=0$ state; without this laser, the scattering rate remains fairly low, which limits the trapping force.
Our numerical simulations predict that this scheme has high compression, realizing a size of $\sim$$50~\text{\textmu m}$, and a temperature of $\sim$$80~\text{\textmu K}$. 
The specific parameters for this proposed 1+2 YO scheme are are $I = 41~\text{mW/cm}^2$ (split $28\%/16\%/28\%/28\%$ among the (I)/(II)/(IIIa)/(IIIb) components), $\Delta = 5.0~\text{MHz}$, $\delta_\text{II} = 3.0~\text{MHz}$, $\delta_\text{IIIa} = -0.6~\text{MHz}$, $\delta_\text{IIIb} = 0.6~\text{MHz}$, and $B = 50~\text{G/cm}$.

\section{Trapping mechanism in a toy model}
\begin{figure}[b!]
    \centering
    \includegraphics[width=1\linewidth]{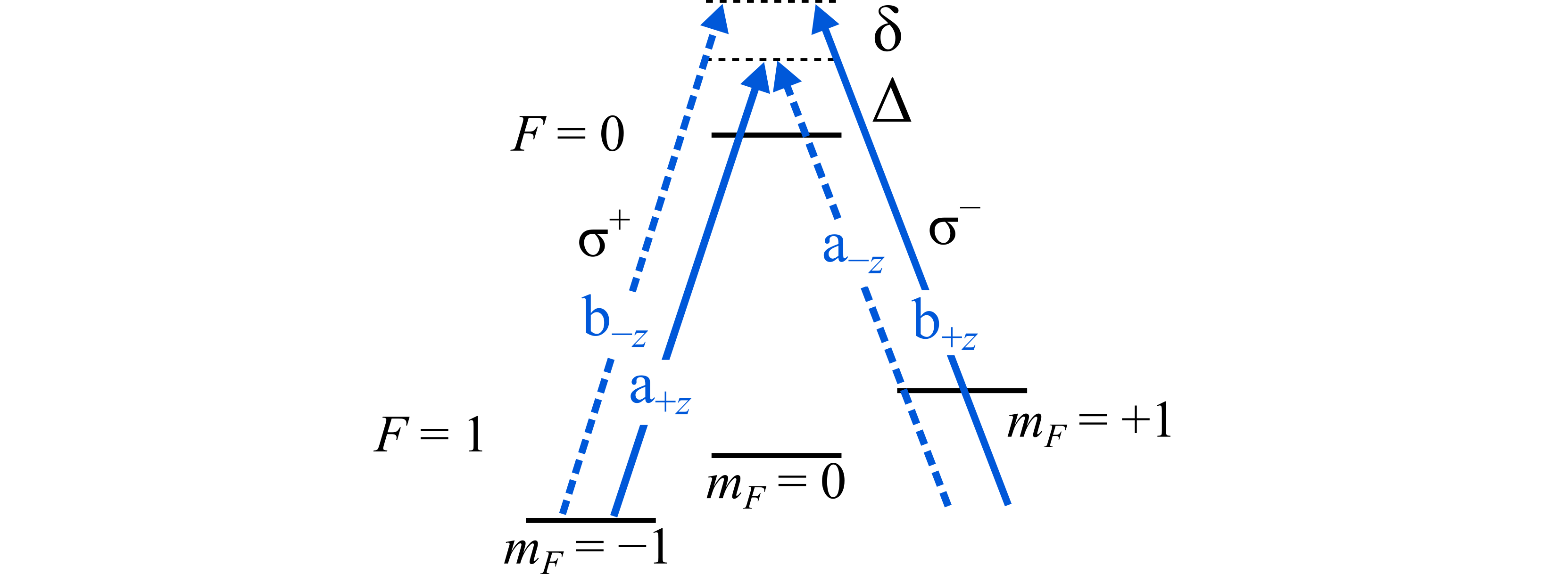}
    \caption{Blue MOT toy model for a $F=1 \to F'=0$ transition, with the $z$ MOT beams shown in arrows. The ingoing MOT beam contains the frequency components $\text{a}_{+z}$ and $\text{b}_{+z}$ (solid lines), while the retroreflected MOT beam contains the frequency components $\text{a}_{-z}$ and $\text{b}_{-z}$ (dashed lines).}
    \label{fig:toy}
\end{figure}

The data presented in the main text clearly demonstrates that addressing $\ket{J=3/2}$ with two closely-spaced frequency components of opposite polarizations leads to a substantial enhancement in the trapping force, which facilitates faster and more effective compression within the MOT. Here we provide preliminary analysis of the mechanism proposed in the main text by considering a blue MOT for a toy model with a single type-two transition, $\ket{F=1} \to \ket{F'=0}$ (Fig.~\ref{fig:toy}).

In the toy model, the MOT beam along the $+\hat{z}$ direction contains two frequency components, $\text{a}_{+z}$ and $\text{b}_{+z}$, with frequencies $\Delta$ and $\Delta + \delta$, and with retroreflections $\text{a}_{-z}$ and $\text{b}_{-z}$ going along $-\hat{z}$. (The $x$ and $y$ MOT beams are defined similarly.) The $\text{b}_{+z}$ component, combined with the retroreflection $\text{a}_{-z}$, makes a slowly moving standing wave (``walking wave'') with $\sigma^-$ polarization. The velocity of the walking wave is $$v_\text{walk} \approx \frac{\delta}{2f_0} c,$$
where $f_0$ is the transition frequency. Similarly, the $\text{a}_{+z}$ and $\text{b}_{-z}$ components form a walking wave with $\sigma^+$ polarization, moving with velocity $-v_\text{walk}$. 
For the $1.75~\text{MHz}$ splitting in the 1+2 scheme of CaOH, $v_\text{walk}$ = 0.55 m/s.

A molecule at rest sitting at a position $z > 0$ (where $B_z > 0$) experiences a Zeeman shift which tunes the blue-detuned MOT light closer to resonance for the $\ket{F=0, m_F=-1} \to \ket{F'=0, m_F' = 0}$ transition, and further away from resonance for the $\ket{F=1, m_F=+1} \to \ket{F'=0, m_F' = 0}$ transition. As a result, the molecule experiences a large AC Stark shift from the $\sigma^+$ walking wave if it is in $\ket{m_F=-1}$, but sees little or no AC Stark shift if it is in $\ket{m_F=+1}$ or   $\ket{m_F=0}$~\cite{devlin2016three}.

In the reference frame comoving with the $\sigma^+$ walking wave, the molecule is travelling with velocity $v = v_\text{walk}$ in a standing wave, and experiences Sisyphus-type cooling: the molecule in $\ket{m_F=-1}$ moves up the standing wave potential hill, gets optically pumped into $\ket{m_F=0,1}$ as it approaches the top of the hill where the optical pumping rate is highest due to the high intensity~\cite{devlin2016three}.
As it nears a potential trough, the molecule can return to $\ket{m_F=-1}$ by either being non-adiabatically remixed by transverse magnetic fields \cite{emile1993magnetically} or by absorbing from a transverse laser beam, allowing it to continue the hill-climbing cycle and decelerate in the moving frame.
% As it nears a potential trough, the molecule can return to $\ket{m_F=-1}$ through \cite{graymolasses, emile1993magnetically} or by absorbing from a transverse laser beam, allowing it to continue the hill-climbing cycle and decelerate in the moving frame.

\begin{figure}[t!]
    \centering
    \includegraphics[width=1\linewidth]{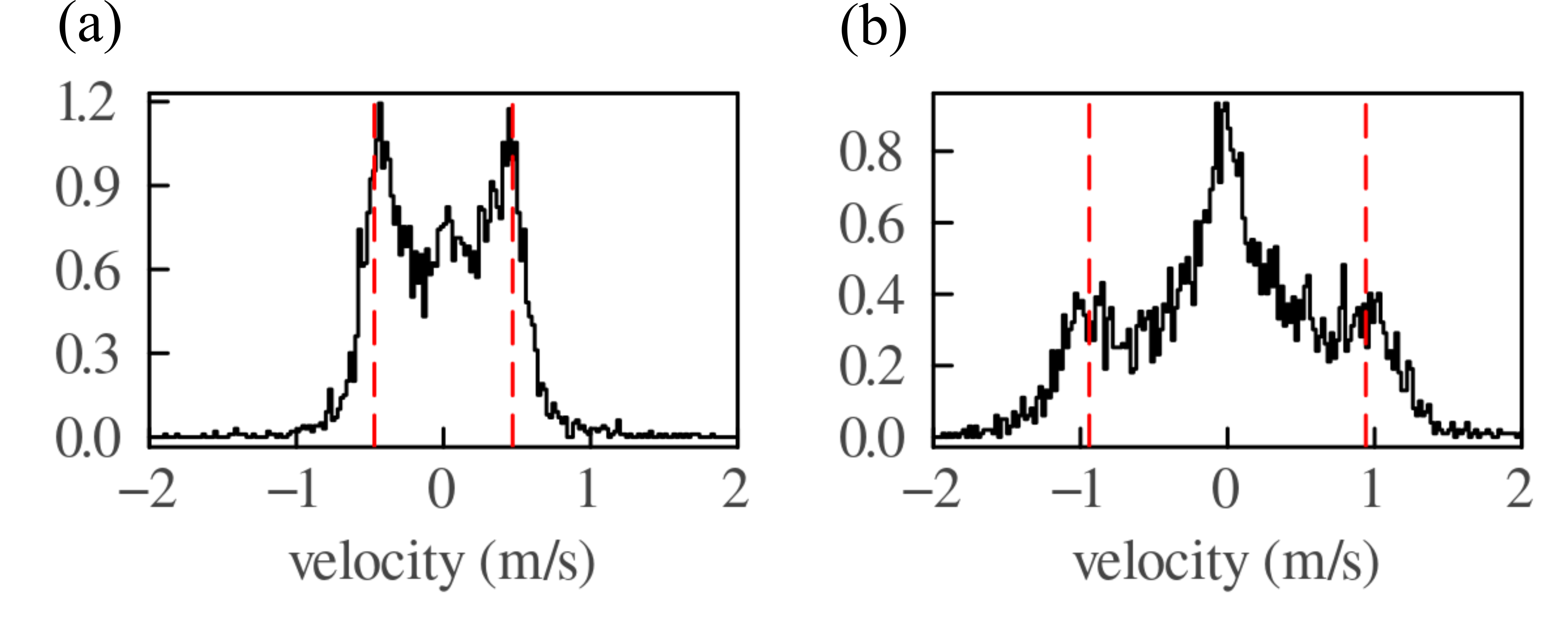}
    \caption{Velocity distribution of the toy model molecules during the compression for (a) $\delta = 1.5~\text{MHz}$ and (b) $\delta = 3~\text{MHz}$. The red vertical lines mark the corresponding velocity of the walking waves $v_\text{walk}$ of (a) $0.47~\text{m/s}$ and (b) $0.94~\text{m/s}$. Parameters used for both simulations are $B'_z$ = 20 G/cm, $\Delta$ = 10 MHz, $s_a$ = $s_b$ = 6. Both distributions are taken at $t = 0.6~\text{ms}$, which is during the compression, before reaching equilibrium. }
    \label{fig:hist}
\end{figure}

The cooling force in the moving frame translates to a pushing force in the lab frame, accelerating the molecule toward the center (since the $\sigma^+$ walking wave is moving with negative velocity $-v_\text{walk}$). As the molecule nears the MOT center, the magnetic field becomes smaller, and the trapping mechanism becomes weaker since the molecule interacts equally strongly with both walking waves due to the vanishing Zeeman shift. At the center of the MOT, gray-molasses cooling in the lab frame in the $\sigma^+ \sigma^-$ polarization gradient becomes dominant over the pushing force of the walking waves. Consequently, the molecule decelerates and remains near the center.

\begin{figure}[t!]
    \centering
    \includegraphics[width=1\linewidth]{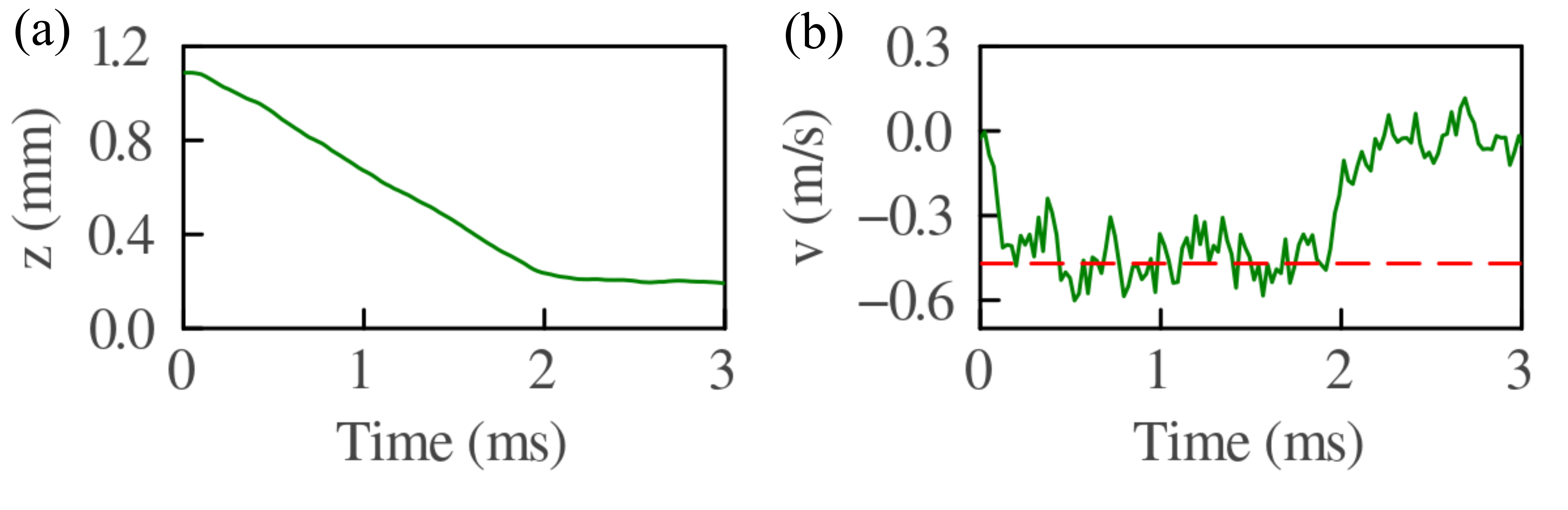}
    \caption{Position and velocity along the $\hat{z}$ direction for a single particle trajectory for $\delta = 1.5~\text{MHz}$. The red dashed line indicates $v_\text{walk} = -0.47 $ m/s.}
    \label{fig:trajectory_example}
\end{figure}

We perform numerical simulations for the toy model to investigate if the proposed trapping mechanism takes place for this case. We set the Land\'e g-factor of the $F=1$ state to be $g = +0.83$, which is the same g-factor as that of the $|J=3/2, F=1\rangle$ state in CaOH. According to our proposed trapping mechanism, we anticipate that molecules displaced from the MOT center will be accelerated to the velocity of one of the walking waves, $\pm v_\text{walk}$. This is exactly what we observe in simulations. Fig.~\ref{fig:hist} illustrates the velocity distribution during compression for two different laser schemes with splitting $\delta = 1.5~\text{MHz}$ and $\delta = 3~\text{MHz}$, corresponding to walking wave speed speed $v_\text{walk} = 0.47~\text{m/s}$ and $v_\text{walk} = 0.94~\text{m/s}$, respectively. In both simulations, the saturation parameter $s = I/I_\text{sat}$ is set to 6 for all four frequency components $a_{+z}$, $a_{-z}$, $b_{+z}$, and $b_{-z}$. Here $I_\text{sat} = \pi h c \Gamma / 3\lambda^3$ is the saturation intensity for a two-level system with linewidth $\Gamma$ and wavelength $\lambda$.
The population accumulation around $\pm v_\text{walk}$ in both cases indicates that the molecules fly towards the center with same velocity as the walking wave. An example trajectory showcasing this process is plotted in Fig.~\ref{fig:trajectory_example}.

It may be possible that ``cooling effects'' in a moving frame could also play a role for the behavior observed with respect to two-photon detuning for the 1+1 schemes (sec. III). Consider for example a $\Lambda$-type system realized by two counterpropagating laser beams with nonzero two-photon detuning $\delta$. Such a system has a velocity-selective dark state at a nonzero velocity $v = \delta/2k$~\cite{cheuk2018lambda}. If molecules in the blue MOT interact preferentially with certain MOT beams as a function of position due to Zeeman shifts, the presence of such dark states could lead molecules to ``cool'' towards certain velocities that are relatively ``dark'' in different locations of the MOT. In the lab frame, the cooling would induce pushing and, potentially, a restoring force. Such a force would indeed flip its sign when $\delta$ changes sign, as observed for the 1+1 (A) scheme, since the sign of the ``dark'' velocities reverses around $\delta = 0$. Further investigation is required to determine if such effects take place in molecular MOTs.

\bibliography{CaOH_BlueMOT_References}

\newpage